\documentstyle[aps,12pt,epsf]{revtex}
\def\la{\mathrel{\mathpalette\fun <}}
\def\ga{\mathrel{\mathpalette\fun >}}
\def\fun#1#2{\lower3.6pt\vbox{\baselineskip0pt\lineskip.9pt
  \ialign{$\mathsurround=0pt#1\hfil##\hfil$\crcr#2\crcr\sim\crcr}}}

\def\nualpha{{\nu_\alpha}}
\def\nualphabar{{\bar{\nu}_\alpha}}
\def\nubeta{{\nu_\beta}}
\def\nubetabar{{\bar{\nu}_\beta}}

\def\mus{$\nu_\mu\leftrightarrow\nu_s\ $}
\def\taus{$\nu_\tau\leftrightarrow\nu_s\ $}
\def\mutau{$\nu_\mu\leftrightarrow\nu_\tau\ $}
\begin{document}

\draft
\title{Big Bang Nucleosynthesis and Active-Sterile Neutrino Mixing:
Evidence for Maximal $\nu_\mu\leftrightarrow\nu_\tau$
Mixing in Super Kamiokande?}

\author{Xiangdong~Shi and George~M.~Fuller}
\address{Department of Physics, University of California,
San Diego, La Jolla, California 92093-0319}

\maketitle

\begin{abstract}
We discuss Big Bang Nucleosynthesis constraints on
maximal $\nu_\mu\leftrightarrow\nu_s$ mixing.
Vacuum $\nu_\mu\leftrightarrow\nu_s$ oscillation has
been proposed as one possible explanation of
the Super Kamiokande atmospheric neutrino data.
Based on the most recent primordial abundance
measurements, we find that the effective number of
neutrino species for Big Bang Nucleosynthesis (BBN)
is $N_\nu\la 3.3$.
Assuming that all three active neutrinos are light
(with masses $\ll 1$ MeV), we examine BBN
constraints on $\nu_\mu\leftrightarrow\nu_s$ mixing
in two scenarios:
(1) a negligible lepton asymmetry (the standard picture);
(2) the presence of a large lepton asymmetry which has
resulted from an amplification by $\nu_\tau\leftrightarrow\nu_{s^\prime}$
mixing ($\nu_{s^\prime}$ being $\nu_s$ or another sterile
neutrino species). The latter scenario has been proposed
recently to reconcile the BBN constraints and
large-angle $\nu_\mu\leftrightarrow\nu_s$ mixing.
We find that the large-angle $\nu_\mu\leftrightarrow\nu_s$
mixing in the first scenario, which would yield $N_\nu\approx 4$,
is ruled out as an explanation of the Super Kamiokande data.
It is conceivably possible for the $\nu_\mu\leftrightarrow\nu_s$ 
solution to evade BBN bounds in the second scenario, but
only if 200 eV$^2\la m^2_{\nu_\tau}-m^2_{\nu_{s^\prime}}\la 10^4$
eV$^2$ is satisfied, and if $\nu_\tau$ decays non-radiatively with
a lifetime $\la 10^3$ years.  This mass-squared difference
implies 15 eV$\la m_{\nu_\tau}\la 100$ eV if $\nu_{s^\prime}$
is much lighter than $\nu_\tau$. 
We conclude that maximal (or near maximal)
$\nu_\mu\leftrightarrow\nu_\tau$ mixing is
a more likely explanation of the Super Kamiokande data.

\end{abstract}
\bigskip


\newpage
\section{Introduction}
The recent Super Kamiokande (SuperK) data on atmospheric
neutrinos show strong evidence for
neutrino oscillation in a channel which involves muon
neutrinos and another neutrino species, and with vacuum
mass-squared difference $3\times 10^{-4}$ eV$^2<\delta m^2
<7\times 10^{-3}$ eV$^2$ and vacuum mixing angle
satisfying $0.8<\sin^2 2\theta\le 1$\cite{SuperK,sterile}.
Measurements of the muon-like events and limits on the 
electron-like events preclude mixing between $\nu_\mu$ and $\nu_e$
as a significant channel for $\nu_\mu$ reduction\cite{SuperK}.
A $\nu_\mu\leftrightarrow\nu_e$ mixing with these parameters
would also yield a uniform suppression of the solar
neutrino flux by roughly a factor $\sim 2$ across the entire
solar neutrino energy spectrum. However, such an energy
independent suppression contradicts what is implied from
the combined solar neutrino data \cite{Solar}.
This leaves two maximal or near maximal vacuum neutrino
oscillation channels as possible avenues for solution:
\mutau mixing; or \mus mixing, where $\nu_s$ is a sterile
neutrino (e.g., a singlet under the
SU(3)$\times$ SU(2)$\times$ U(1) gauge symmetry of the Standard Model).  

However, it has been pointed out by many groups that \mus\ 
mixing with the aforementioned parameters would violate the
constraints from Big Bang Nucleosynthesis (BBN) if all three
active neutrinos are light compared to 1 MeV. This violation
results from $\nu_\mu\leftrightarrow\nu_s$ mixing essentially
bringing an extra $\nu_s$ into chemical equilibrium with the
three known species at the BBN epoch.
In turn, these extra degrees of freedom imply a larger
expansion rate and so would yield a primordial $^4$He
abundance that is too high to accommodate
observationally-determined abundance bounds
\cite{Dolgov,Enqvist1,Cline,Shi1}.

There has been some confusion on this result lately
due to a larger systematic uncertainty in the measured
primordial $^4$He abundance as argued by some groups,
and due to recent measurements of the primordial deuterium
abundance.  Furthermore, the BBN constraint has been
obtained by assuming a negligible lepton number asymmetry 
($\la 10^{-7}$), which has been shown to be an invalid
assumption under circumstances where the lepton asymmetries
in the neutrino sector can be amplified by active-sterile 
neutrino transformations\cite{Foot1,Shi2}. It has been 
argued that a $\nu_\tau$ mixing with a lighter sterile 
$\nu_{s^\prime}$ may amplify an initially negligible lepton
asymmetry to a large level that either is sufficient to
suppress a subsequent $\nu_\mu\leftrightarrow\nu_s$ mixing 
\cite{Foot1}, or can be partly converted to a positive chemical
potential in the $\nu_e\bar\nu_e$ sector to lower the effective
$N_\nu$\cite{Foot3}. Both may conceivably lead to an evasion of
the BBN bound on the $\nu_\mu\leftrightarrow\nu_s$ mixing parameters.

It is traditional in BBN studies to parametrize both the $^4$He
yield and the expansion rate of the universe at the BBN epoch
in terms of an effective number of relativistic neutrino flavors,
$N_\nu$.  In the context of neutrino oscillation discussions
this convention is potentially confusing and misleading, since
efficient neutrino matter-enhanced transformation could result
in both a larger neutrino energy density and a lower $^4$He
yield (by way of a $\nu_e\bar\nu_e$ asymmetry)
than in the standard picture!  Here we will also employ
$N_\nu$ in the traditional manner to facilitate comparison
with the results of previous work.  However, we will attempt
to point out what the true underlying picture is in every case.

In this paper we intend to clarify the bound from BBN in
light of the latest development in the primordial $^4$He
and D abundance measurements, and critique
the possibility that the \mus\ mixing parameters required
to explain the SuperK data might survive the BBN
bound because of an amplified lepton asymmetry.
We implicitly assume that all three active neutrinos
are light, being relativistic at the BBN epoch.

\section{Formalism}
Here we briefly outline the formalism for active-sterile
neutrino transformation used in this paper.
A detailed treatment of active-sterile neutrino mixings in the
BBN epoch can be found in ref.~\cite{Enqvist1} or \cite{Shi1}.

We adopt units in which $\hbar=c=k=1$. We denote by $n_i$
the number density of a particle $i$ relative to its equilibrium
value at a temperature $T$. These equilibrium values are
$2\zeta (3)T^3/\pi ^2$ for photons, $3\zeta (3)T^3/2\pi ^2$ for
electrons/positrons, and $3\zeta (3)T^3/4\pi ^2$ for each
neutrino species.
We describe the $\nu_\alpha\leftrightarrow\nu_s$
($\alpha=e$, $\mu$ or $\tau$) transformation channel
at the BBN epoch with the following two differential equations:
\begin{equation}
{{\rm d}P_0\over {\rm d}t}=\sum_{i=e,\nu_\beta; \beta\neq\alpha}
\bigl\langle\Gamma(\nualpha\nualphabar\rightarrow i\bar i)\bigr\rangle
(n_i n_{\bar i}-n_\nualpha n_{\nualphabar});
\label{totnum}
\end{equation}
\begin{equation}
{{\rm d}{\bf P}\over {\rm d}t}
={\bf V}\times{\bf P}+{{\rm d}P_0\over {\rm d}t}{\bf\hat z}-D{\bf P_\bot}.
\label{master}
\end{equation}
In the equations, $P_0$ denotes the total number density of the mixture
of active and sterile neutrino species,
$P_0=n_{\nu_\alpha}+n_{\nu_s}$, and {\bf P} denotes the composition
of the mixture.  In particular, $P_z=n_{\nu_\alpha}-n_{\nu_s}$.  The
other component of {\bf P}, {\bf P}$_\bot=P_x{\bf\hat x}+P_y{\bf\hat y}$,
is an indication of the phase coherence of the oscillation channel.
Quantities $\langle\Gamma\rangle$ are thermally averaged reaction rates.
The vector {\bf V} represents the frequency and the axis of the oscillation
in {\bf P}-space, and the $D$-term represents the damping of
{\bf P}$_\bot$ due to neutrino interactions. This latter term
constantly acts to reduce a mixed neutrino state into flavor 
eigenstates of $\nu_\alpha$ and $\nu_s$.

At the epoch of BBN,
\begin{equation}
V_x={\delta m^2\over 2E}\sin 2\theta,\quad
V_y=0,\quad
V_z=-{\delta m^2\over 2E}\cos 2\theta+V_\alpha^L+V_\alpha^T,
\label{Vcomponent}
\end{equation}
where $\delta m^2=m^2_{\nu_s}-m^2_{\nu_\alpha}$
and $\theta$ are the usual vacuum mixing parameters
of mass-squared difference and effective vacuum 2$\times$2 mixing
angle ($\delta m^2>0$ if $\nu_s$ is heavier than $\nu_\alpha$), and
$E$ is the energy of the neutrinos. $V_\alpha^L$ is the contribution of
the MSW matter-enhanced effect driven by asymmetries
in the background plasma \cite{Raffelt}, and is approximately
\begin{equation}
V_\alpha^L\approx 0.13 G_FT^3\Bigl[8L_0/3
                  +2(n_{\nu_\alpha} -n_{\nualphabar})+\sum_
                  {\nubeta\neq\nualpha}(n_\nubeta-n_\nubetabar)\Bigr]
\label{VL}
\end{equation}
where $G_F$ is the Fermi constant, $T$ is the plasma temperature,
and $L_0$ represents the contributions from the baryonic asymmetry
and electron-positron asymmetry, $\sim 10^{-9}$.
$V_\alpha^T$ is the contribution due to finite temperature
neutrino mass renormalization effects \cite{Raffelt},
and for $\alpha=\mu$ or $\tau$, is
\begin{equation}
V_\alpha^T \approx -11(n_{\nu_\alpha}+n_{\bar\nu_\alpha})G_F^2ET^4. 
\label{VT}
\end{equation}
Meanwhile, for $\alpha=\mu$ or $\tau$, we also have
\begin{equation}
D\sim 0.5G_F^2T^4E.
\label{Dterm}
\end{equation}
It has been shown that adopting the average neutrino
energy $E\approx 3.151T$ in Eq.~(\ref{Vcomponent})--(\ref{Dterm})
and thermally averaging interaction rates give a fairly good
description of active-sterile neutrino transformations in BBN
\cite{McKeller}.

The same equations (\ref{totnum}) through (\ref{VT}) also
apply to the $\bar\nu_\alpha\leftrightarrow\bar\nu_s$
mixing if we switch $n_{\nu_\alpha}$ ($n_{\nu_\beta}$)
and $n_{\bar\nu_\alpha}$ ($n_{\bar\nu_\beta}$) in the 
expression of $V_\alpha^L$. Here we add an overbar
on top of all variables associated with the
$\nu_\alpha\leftrightarrow\nu_s$ system to 
denote the corresponding variables in
the $\bar\nu_\alpha\leftrightarrow\bar\nu_s$ system.

The initial condition of the universe is taken to be
a pure $\nu_\alpha\bar\nu_\alpha$ system 
(i.e., $P_0=\bar P_0=P_z=\bar P_z=1$ and $P_x=\bar P_x=P_y
=\bar P_y=0$) at an initial temperature
\begin{equation}
T_{\rm init}\gg 
15\left\vert{\delta m^2\cos 2\theta\over {\rm eV}^2}\right\vert^{1/6}.
\label{Tinit}
\end{equation}
This temperature is taken to be high enough that
$V_\alpha^T$ dominates over $\delta m^2/2E$, guaranteeing that
any neutrino transformation is severely suppressed.
(Note however, that $T_{\rm init}$ has to
be below the temperature of the Quark-Hardron phase transition,
$\sim 150$ MeV \cite{Malaney}.)

Barring the existence
of a large initial lepton asymmetry generated by unknown
processes at $T\ga 150$ MeV, we will assume a negligible initial
lepton asymmetry $L$ (e.g., at the same level as $L_0$).
Here a significant $L$ can be generated only
by active-sterile neutrino transformation during the BBN epoch.

\section{BBN Bound When Lepton Asymmetry Is Negligible}

The primordial $^4$He yield $Y$ in standard BBN is principally
a function of the expansion rate of the universe, as this
determines the temperature of weak freeze-out and, hence,
the overall neutron-to-proton ratio. Secondarily, the $^4$He
abundance also depends on the baryon density $N_{\rm B}$
(usually expressed as $\eta\equiv N_{\rm B}/N_\gamma$),
albeit only weakly.  Therefore, given an $\eta$ inferred
from the primordial deuterium abundance D/H (which depends
sensitively on $\eta$), a bound on $N_\nu$ in the standard
BBN picture can be obtained based on the measured
primordial $^4$He abundance\cite{Schramm}.
The same bound can be used to constrain active-sterile
neutrino mixings, because such mixings can produce extra
neutrino energy density.  However, one should use caution
here because these scenarios could also result in a
larger number of $\nu_e$'s over $\bar\nu_e$'s (a positive
$\nu_e\bar\nu_e$ asymmetry), which could lead to a
reduced neutron-to-proton ratio and, hence, a lower $Y$.

Previous bounds on these mixing channels have
been calculated assuming a small $\eta$ 
($\approx 3\times 10^{-10}$) because an
accurate measurement of the primordial D/H was lacking.
It is now, however, established that the primordial
D/H lies at the low end of the range previously thought:
D/H$\approx 3.4\pm 0.3\,\times 10^{-5}$\cite{Tytler1}.
This low value infers a higher $\eta\approx
5.2\pm 0.3\times 10^{-10}$,
which predicts a slightly higher primordial $^4$He
abundance $Y$ for the standard picture $N_\nu=3$.
The upper limit on $N_\nu$ based on the measured $Y$
should then become tighter\cite{Cardall}.

However, complicating this issue is the debate over
the systematic uncertainties in the observationally-inferred
$Y$.  Based on 62 low metallicity HII regions, Olive {\sl et al.}
\cite{Olive} estimated  $Y=0.234\,\pm 0.002\,(stat.)\,\pm 0.005\,(sys.)$;
or $Y=0.237\,\pm 0.003\,(stat.)\,\pm 0.005\,(sys.)$,
when the controversial object I Zw 18 (also the most 
metal-deficient object of the sample) is ignored.
Based on another sample which partially overlaps with
that of Olive {\sl et al.}, Izotov and Thuan \cite{Izotov} 
have claimed $Y=0.244\pm 0.002$, partly because they deduced
a higher $Y$ for I Zw 18.
Disagreement aside, the two values do not seriously
contradict each other given the systematic uncertainty
$\sim 0.005$ of the estimates.  With the systematic uncertainty
in mind, we can still set a conservative upper limit $Y<0.25$ for
the purpose of constraining the neutrino oscillation physics at
the BBN epoch.  A higher $Y$ would not only be inconsistent with
both estimates, it could also possibly be inconsistent with the 
morphology of the Horizontal Branch stars in globular clusters\cite{Dorman}.

Thus, given D/H$\approx 3.4\pm 0.4\,\times 10^{-5}$ and $Y<0.25$, we can
set a rather conservative bound on $N_\nu$ at the $95\%$ C.L.,
\begin{equation}
N_\nu\la 3.3.
\end{equation}

This bound can be used to constrain the active-sterile neutrino
transformations in the BBN epoch, but it is more precise and
convenient to calculate directly the helium yield of
neutrino oscillations, because it is ultimately the helium
mass fraction that is compared to observations.
We took $\eta=4.6\times 10^{-10}$ and
recalculated the $^4$He yield in the presence of
active-sterile neutrino mixing. (Our calculations are similar
to those in Shi, Schramm and Fields\cite{Shi1}.)
Figure 1 shows the result.  A parametrization of the bound on the
$\nu_\mu\leftrightarrow\nu_s$ mixing can be obtained:
\begin{equation}
\vert\delta m^2\vert\sin^4 2\theta\la 10^{-5}\,(10^{-7})\,{\rm eV}^2\
{\rm for}\ \delta m^2>(<)0.
\label{BBNbound}
\end{equation}
The \mus\ mixing solution to the SuperK atmospheric
neutrino data is clearly ruled out by BBN if lepton asymmetry
is negligible ($L<10^{-7}$ in this case).  To put this result
in perspective, our results suggest that a $\nu_\mu\leftrightarrow\nu_s$
mixing solution to the SuperK data in this scenario
would imply $N_\nu=3.9$.

\section{BBN Bound in the Presence of an Amplifed Lepton Asymmetry}

A resonant $\nu_\alpha\leftrightarrow\nu_s$ transformation
can amplify the lepton asymmetry in the $\nu_\alpha$ sector
to a level
\begin{equation}
L_{\nu_\alpha}\equiv{3(n_{\nu_\alpha}-n_{\bar\nu_\alpha})\over 8}
\sim \pm{\vert\delta m^2/{\rm eV}\vert\over 10 T_6^4},
\label{asymptotic}
\end{equation}
where $T_6=T/{\rm MeV}$\cite{Shi2}.
Generation of this lepton number does
not necessarily force $\nu_s$ into chemical equilibrium
if the mixing is small enough to satisfy the constraints in
Fig.~1\cite{Shi2}.

To calculate the amplification of the lepton asymmetry by
$\nu_\alpha\leftrightarrow\nu_s$ mixing, we take the initial
lepton number to be negligibly small.  Note that in this
limit $L_{\nu_\alpha}=3(P_z-\bar P_z)/16$.
From Eq.~(\ref{master}) and its anti-neutrino counterpart,
$L_{\nu_\alpha}$ satisfies the approximate equation
\begin{equation}
{{\rm d}L_{\nu_\alpha}\over {\rm d}t}
\approx{DV_x^2\over V_x^2+[V_{0}-\beta(2L_{\nu_\alpha}+L_0)]^2}
        \left\{{3V_{0}\beta (2L_{\nu_\alpha}+L_0)P_z\over
 4 \{V_x^2+[V_{0}+\beta (2L_{\nu_\alpha}+L_0)]^2\}L_{\nu_\alpha}}
-1\right\}L_{\nu_\alpha},
\label{mastereq}
\end{equation}
which is valid in the adiabatic limit,
$\vert{\bf V}\vert\gg\vert{\rm d}{\bf V}/{\rm d}t
\vert/\vert{\bf V}\vert$.
In Eq.~(\ref{mastereq}) we use $V_0=(V_z+\bar V_z)/2$,
and $\beta=(V_z-\bar V_z)/2(2L_{\nu_\alpha}+L_0)
\approx 0.35G_FT^3$.  We have also assumed
$P_z\sim 1$, because the $\nu_\alpha\leftrightarrow\nu_{s}$
mixing would violate the BBN bound in Fig. 1 if $P_z\ll 1$.

Eq.~(\ref{mastereq}) is a damping equation for
$2L_{\nu_\alpha}+L_0$, unless $V_0>0$.
Here $V_0>0$ if $\delta m^2<0$
(i.e., a heavier $\nu_\alpha$) and if the temperature
$T$ drops below the resonant temperature 
for $\nu_\alpha\leftrightarrow\nu_s$ (see the Appendix),
\begin{equation}
T_{\rm res}\approx 22\,
\left\vert{\delta m^2\over 1{\rm eV}^2}\right\vert^{1/6}{\rm MeV}
\quad\quad {\rm for\ } \alpha=\mu,\tau.
\label{Tres22}
\end{equation}
For $T>T_{\rm res}$,  $2L_{\nu_\alpha}+L_0$ will be driven toward zero.
In this case $V_{\nu_\alpha}^L$ approaches zero,
unless the initial asymmetry $L_0$ (whose definition
may be expanded here to also include the neutrino asymmetry
generated by other active-sterile
neutrino mixings) is too large for the damping to be efficient.
Efficient damping occurs when the generated lepton asymmetry
in other neutrino species satisfies
\begin{equation}
\vert L\vert\la 10^{-4}\vert\delta m^2\sin 2\theta\vert^{1/2},
\end{equation}
so that $({\rm d}L_{\nu_\alpha}/{\rm d}t)/L_{\nu_\alpha}\,<\,H$,
where $H$ is the Hubble expansion rate.

Eq.~(\ref{master}) and its anti-neutrino counterpart are
non-linear equations. It has two attractors
which give the asymptotic values of Eq.~(\ref{asymptotic}).
The asymmetry $L_{\nu_\alpha}$ calculated through
numerical integrations of these equations exhibits chaotic
behavior for $T$ just below $T_{\rm res}$\cite{Shi2}.  In this phase
$L_{\nu_\alpha}$ oscillates chaotically between
the two attractors.  For sufficiently low $T$, $L_{\nu_\alpha}$
approaches asymptotically one of the two attractors.  It is in
this asymptotic phase when the adiabatic condition is satisfied
and Eq.~(\ref{mastereq}) applies.

It is conceivable that a $\nu_\tau\leftrightarrow\nu_{s^\prime}$
mixing satisfying
\begin{equation}
\left\vert{\delta {m^\prime}^2\over{\rm eV}^2}\right\vert^{1/6}
\sin^22\theta\ga 10^{-11}.
\end{equation}
and the the BBN bound Eq.~(\ref{BBNbound}),
can generate $\vert L_{\nu_\tau}\vert\gg 10^{-7}$\cite{Shi2,Foot2},
which in turn suppresses the bad BBN effects of
$\nu_\mu\leftrightarrow\nu_s$ transformation with
$\delta m^2\sim 10^{-3}$ eV$^2$ and $\sin^2 2\theta\sim 1$,
and so relaxes the BBN bound on the $\nu_\mu\leftrightarrow\nu_{s}$ mixing.
(Here $\nu_{s^\prime}$ and $\nu_s$ could be in principle
the same species but are distinguished here for clarity.)
But as we will demonstrate below, while this is not impossible,
there are severe constraints on the required
$\nu_\tau\leftrightarrow\nu_{s^\prime}$ mixing
parameters for this loophole to be realized.

To distinguish between the $\nu_\tau\leftrightarrow\nu_{s^\prime}$
transformation and the \mus transformation,
all quantities applying to the former process are marked with
an prime. To suppress the transformation due to a near maximal
\mus mixing, the lepton asymmetry has to be amplified early by
the $\nu_\tau\leftrightarrow\nu_{s^\prime}$ transformation
channel before any significant \mus transformation can occur.
This requires the $\nu_s$ production rate at the
$\nu_\tau\leftrightarrow\nu_{s^\prime}$ resonance temperature
$T_{\rm res}^\prime$ to satisfy
\begin{equation}
D\left({\delta m^2\sin 2\theta\over V_\mu^T}\right)^2\,<H,
\label{prod}
\end{equation}
with $H=5.5T^2/m_{\rm pl}$. Here $m_{\rm pl}\approx
1.22\times 10^{28}$ eV is the Planck mass.
Employing $T_{\rm res}^\prime$ and $V_\mu^T$
we obtain from Eq.~(\ref{prod}):
\begin{equation}
\vert\delta {m^\prime}^2\vert > 10\,\vert\delta m^2\sin 2\theta\vert^{4/3}
\sim 10^{-2}\,{\rm eV}^2.
\end{equation}

Once the lepton asymmetry is amplified, the \mus mixing will have
\begin{equation}
V_x=-{\delta m^2\over 2E}\sin 2\theta,\quad 
V_z=-{\delta m^2\over 2E}\cos 2\theta+V_\mu^T\pm V_\mu^L
\approx\,V_\mu^T\pm \beta(2L_{\nu_\mu}+L_{\nu_\tau}),
\label{ptl}
\end{equation}
where \lq\lq +\rq\rq\ (\lq\lq $-$\rq\rq) applies to the \mus
($\bar\nu_\mu\leftrightarrow\bar\nu_s$) transformation channel.
If $L_{\nu_\tau}$ is positive (negative), the \mus 
($\bar\nu_\mu\leftrightarrow\bar\nu_s$) channel will experience
a resonance, regardless of the sign of $\delta m^2$, at a neutrino energy
\begin{equation}
E_{\rm res}\approx {\beta\vert L_{\nu_\tau}\vert\over V_\mu^T/E}
\approx {\vert L_{\nu_\tau}\vert\over 63G_FT}
\end{equation}
(assuming initially $L_{\nu_\mu}=0$).
The resonance moves from lower energies to higher energies in the
mu neutrino spectrum as $\vert L_{\nu_\tau}\vert$
grows. At the same time, the $\bar\nu_\mu\leftrightarrow\bar\nu_s$
(\mus) transformation channel is suppressed by $L_{\nu_\tau}$ as
a consequence of having a even larger $\vert V_z\vert$.
The net result is a newly generated $L_{\nu_\mu}$ with a sign
opposite to that of $L_{\nu_\tau}$.
If this $L_{\nu_\mu}$ becomes as large as $\sim -L_{\nu_\tau}/2$,
the $\nu_\mu\leftrightarrow\nu_s$ mixing will be unhindered and
will eventually bring $\nu_s$ into equilibrium.
Based on this consideration, Foot and Volkas \cite{Foot2,Foot4}
argued for a requirement on the \taus mixing
\begin{equation}
\vert\delta {m^\prime}^2\vert\ga 20 \,{\rm eV}^2.
\end{equation}
Here we show that this requirement is too weak.  We find that
a more stringent requirement can be obtained based on the
same consideration at the beginning stage of the
$L_{\nu_\tau}$ growth, when both $L_{\nu_\tau}$ and the
mu neutrino resonance energy $E_{\rm res}$ are very small.
For definitiveness, we assume that $L_{\nu_\tau}>0$
so that both the $\bar\nu_\tau\leftrightarrow\bar\nu_{s^\prime}$
transformation and the $\nu_\mu\leftrightarrow\nu_s$ transformation
encounter resonances.

Each of these two resonances occurs in a finite neutrino energy
bin (resonance width).
Outside the resonance energy bins the effective mixing between 
active neutrinos and sterile neutrinos quickly diminishes and 
becomes negligible. This feature enables a simple semi-analytical
approach to accurately calculate the growth of lepton number 
asymmetries by tracking only neutrinos in resonance regions.
The details of this semi-analytical approach are presented in the
Appendix.

The constraint on $\delta {m^\prime}^2$ results from the
requirement that $\vert L_{\nu_\tau}\vert$ be much larger 
than $\vert L_{\nu_\mu}\vert $ at any temperature as the
$\nu_\mu\leftrightarrow\nu_s$ resonance sweeps through the
entire neutrino energy spectrum. In other words, we must have
\begin{equation}
f(\epsilon_{\rm res})\delta\epsilon_{\rm res}\,R\,
\left\vert{\delta\epsilon_{\rm res}
\over {\rm d}\epsilon_{\rm res}/{\rm d}t}\right\vert
< {4\over 3}\left(L_{\nu_\tau}+2L_{\nu_\mu}\right)
\label{requirement}
\end{equation}
for any $\epsilon_{\rm res}$.
In the equation, $f(\epsilon)=[2/3\zeta(3)]
[\epsilon^2/({\rm e}^\epsilon+1)]$ is the neutrino distribution function. 
$\delta\epsilon_{\rm res}$ is the energy width of the resonance.
$f(\epsilon_{\rm res})\delta\epsilon_{\rm res}$ is therefore the
fraction of mu neutrinos undergoing resonance. $R$ is the resonant
transition rate.
$\delta\epsilon_{\rm res}/\vert{\rm d}\epsilon_{\rm res}/{\rm d}t\vert$
is the duration of the resonance at $\epsilon_{\rm res}$.
The energy width of the resonance depends on whether the 
resonant transition is collision-dominated ($D>V_x$),
or oscillation-dominated ($D<V_x$):
\begin{equation}
\delta\epsilon_{\rm res} \sim \left\{  \begin{array}{ll}
2\left\vert D{\partial\epsilon_{\rm res}/\partial V_z}\right\vert &
\mbox{if $D>V_x$}\\
2\left\vert V_x{\partial\epsilon_{\rm res}/\partial V_z}\right\vert &
\mbox{if $D<V_x$}\end{array}\right..
\label{width}
\end{equation}
Likewise, the resonant transition rate depends on these parameters:
\begin{equation}
R\approx \left\{  \begin{array}{ll}
V_x^2/D & \mbox{if $D>V_x$}\\
V_x     & \mbox{if $D<V_x$}\end{array}
\right..
\label{transitionrate}
\end{equation}

We only consider the case where $\epsilon_{\rm res}\ll 1$, because
this is when $L_{\nu_\tau}$ is in its
initial stage of growth ($L_{\nu_\tau}\ll 10^{-7}$) and is
most easily matched by a competing $L_{\nu_\mu}$.
In this case, $T\approx T^\prime_{\rm res}$ (the temperature at
which $L_{\nu_\tau}$ growth starts), and $f(\epsilon_{\rm res})
\approx \epsilon_{\rm res}^2/1.8$.
We can further rewrite $\vert\dot\epsilon_{\rm res}\vert\equiv
H\epsilon_{\rm res}\vert {\rm d}\ln\epsilon_{\rm res}/{\rm d}\ln T\vert$.
Then with Eq.~(\ref{Tres22}) and the expression for $H$,
Eq.~(\ref{requirement}) becomes
\begin{eqnarray*}
\quad\quad
\begin{array}{llr}
\left\vert{\delta{m^\prime}^2/1\,{\rm eV}^2}\right\vert^{11/6}
>2\times 10^4\epsilon_{\rm res}^{-1}
\left\vert {{\rm d}\ln\epsilon_{\rm res}/{\rm d}\ln T}\right\vert^{-1}
\left\vert {\delta m^2/10^{-3}\,{\rm eV}^2}\right\vert^2 &
\mbox{if $D>V_x$;} \\
\left\vert{\delta{m^\prime}^2/1\,{\rm eV}^2}\right\vert^{17/6}
>10^3\epsilon_{\rm res}^{-3}
\left\vert {{\rm d}\ln\epsilon_{\rm res}/{\rm d}\ln T}\right\vert^{-1}
\left\vert {\delta m^2/10^{-3}\,{\rm eV}^2}\right\vert^3 & 
\mbox{if $D<V_x$.} \end{array}\quad\quad\ 
\begin{array}{r}
\mbox{\hfill (23a)}\\\mbox{\hfill (23b)}\end{array}
\label{main}
\end{eqnarray*}

In the initial stage of rapid $L_{\nu_\tau}$ growth,
$\vert {\rm d}\ln\epsilon_{\rm res}/{\rm d}\ln T\vert$
can be related to the growth rate of $L_{\nu_\tau}$
by $\vert {\rm d}\ln\epsilon_{\rm res}/{\rm d}\ln T\vert
\approx\vert {\rm d}\ln L_{\nu_\tau}/{\rm d}\ln T -2\vert
\approx\vert {\rm d}\ln L_{\nu_\tau}/{\rm d}\ln T\vert$
(with $L_{\nu_\mu}$ safely ignored). Our semi-analytical
calculations show that in the initial exponential stage of
$L_{\nu_\tau}$ growth when $L_{\nu_\tau}$ is $\ll 10^{-7}$,
$\vert {\rm d}\ln L_{\nu_\tau}/{\rm d}\ln T \vert$
is well approximated by
\setcounter{equation}{23}
\begin{equation}
\left\vert {{\rm d}\ln L_{\nu_\tau}\over {\rm d}\ln T }\right\vert
\approx 6\times 10^6\sin2\theta^\prime,
\label{upperbound}
\end{equation}
indendent of $\delta {m^\prime}^2$.  Of course, 
$\sin2\theta^\prime$ here must satisfy the BBN bound
Eq.~(\ref{BBNbound}).

Eq.~(23) and (\ref{upperbound}) show that
the most stringent requirement on $m_{\nu_\tau}^2-m_{\nu_{s^\prime}}^2$
does indeed come not from $\nu_\mu\leftrightarrow\nu_s$ resonances
at $\epsilon_{\rm res}\sim 3$, but from resonances centered
at the smallest possible $\epsilon_{\rm res}$ as long as the
$\nu_\mu$ or $\bar\nu_\mu$
transition probability in that resonance energy bin is $\ll 1$.
This condition, expressed as
\begin{equation}
R\left\vert{\delta\epsilon_{\rm res}\over {\rm d}\epsilon_{\rm res}/{\rm d}t}
\right\vert\la 0.1,
\end{equation}
can be rewritten as
\begin{equation}
\epsilon_{\rm res}\ga
\left\vert{\delta{m^\prime}^2\over 1\,{\rm eV}^2}\right\vert^{-1/2}
\left\vert {{\rm d}\ln\epsilon_{\rm res}\over {\rm d}\ln T}\right\vert^{-1/3}
\left\vert {\delta m^2\over 10^{-3}\,{\rm eV}^2}\right\vert^{2/3},
\label{howsmall}
\end{equation}
regardless of the value of $D/V_x$.
It can be shown that $\epsilon_{\rm res}$ is
in the oscillation-dominated regime if
\begin{equation}
\epsilon_{\rm res}\la 0.25 \left\vert
{\delta m^2\over 10^{-3}\,{\rm eV}^2}\right\vert^{1/2}
\left\vert{\delta{m^\prime}^2\over 1\,{\rm eV}^2}\right\vert^{-1/2}.
\end{equation}
Therefore, the most stringent requirement on
$\delta {m^\prime}^2$ comes from the
oscillation-dominated regime for $\sin^22\theta^\prime\ga 10^{-8}$
(while $\vert {{\rm d}\ln\epsilon_{\rm res}/{\rm d}\ln T}\vert\ga 10^3$),
and from collision-dominated regime for $\sin^22\theta^\prime\la 10^{-8}$
(while $\vert {{\rm d}\ln\epsilon_{\rm res}/{\rm d}\ln T}\vert\la 10^3$).

Combining Eq.~(23), Eq.~(\ref{upperbound})
and Eq.~(\ref{howsmall}) yields a requirement on the
mass-squared-difference necessary to effect suppression
of $\nu_\mu\leftrightarrow\nu_s$ transformation at the
Super Kamiokande level:
\begin{equation}
m_{\nu_\tau}^2-m_{\nu_{s^\prime}}^2\ga \left\{ \begin{array}{ll}
200\left(\vert m_{\nu_\mu}^2-m_{\nu_s}^2\vert /10^{-3}\,{\rm eV}^2\right)
^{3/4}\,{\rm eV}^2 & \mbox{for $\sin^22\theta^\prime\ga 10^{-8}$}\\
\left(\sin 2\theta^\prime\right)^{-1/2}
\left(\vert m_{\nu_\mu}^2-m_{\nu_s}^2\vert /10^{-3}\,{\rm eV}^2\right)
\,{\rm eV}^2 & \mbox{for $\sin^22\theta^\prime\la 10^{-8}$}\end{array}
\right..
\label{masterreq}
\end{equation}
Since $\vert\delta {m^\prime}^2\vert =m^2_{\nu_\tau}-m^2_{\nu_s}$,
Eq.~(\ref{masterreq}) implies that we must have
\begin{equation}
m_{\nu_\tau}\ga 15\, {\rm eV}
\end{equation}
to successfully suppress the $\nu_\mu\leftrightarrow\nu_s$
transformation in BBN.

Tau neutrinos this massive would contribute a fraction
of the critical density $\Omega_\nu\approx
0.5h_{50}^{-2}$ (where $h_{50}$ is the Hubble constant in the units
of 50 km/sec/Mpc) in the form of Hot Dark Matter (HDM) \cite{Kolb}.
When normalized to yield the observed structure today,
a matter-dominated flat ($\Omega_{\rm m}=1$) model universe
with $\Omega_\nu \ga 0.3$ (the remainder being mostly Cold Dark Matter)
yields too few Damped Lyman-$\alpha$ systems (protogalaxies)
at a redshift $z\ga 3$ to accommodate observations\cite{Kauffmann,Klypin}.
Due to the free-streaming of neutrinos,
the Hot Dark Matter component reduces the density
fluctuation amplitudes at the galaxy mass 
scale and causes galaxy-sized structures
to form too late ($z\la 1$).
Even an $\Omega_\nu=0.2$ flat, matter-dominated model may
still be in disagreement with observations\cite{Ma97}.
Therefore, any light stable and weakly-interacting
neutrino more massive than $m_\nu\ga 5h_{50}^2$ eV will have
trouble with structure formation considerations.
Models with these neutrinos cannot be rescued by pushing $h_{50}$
much higher, because both the age of the universe and the observed
structure today require $h_{50}\approx 1$ \cite{Gross}.
If as implied by the cosmic deceleration measurements\cite{Kirshner},
the total matter density in our universe is less than
the critical density, the incompatibility of structure formation
at high $z$ and $\Omega_\nu\ga 0.2$ will become even worse -
the HDM would then comprise a larger fraction of the matter density and
so reduce the density fluctuations at small scales even more effectively.

To simultaneously have $m_{\nu_\tau}\ga 15$ eV and successful structure
formation at high redshifts, the massive tau neutrinos would have to decay
before making an imprint on the cosmic density fluctuation spectrum.
Since sub-horizon-sized density fluctuations (i.e., those affected by
free streaming of neutrinos) were
frozen before the epoch of matter-radiation equality and
grew only after that, the density fluctuations residing
in other Dark Matter components would not be affected by
the density fluctuations in the tau neutrino spatial distribution
so long as these $\nu_\tau$'s decayed before
the matter-radiation equality epoch.
The lifetime of $\nu_\tau$ must then be
\begin{equation}
\tau_{\nu_\tau}\la 10^3\ {\rm year}.
\end{equation}
The decay cannot be radiative, as it would violate the bound 
on neutrino electromagnetic dipole moments \cite{Raffelt2}.

For those \taus mixings with $10^{-2}$ eV$^2\la\vert\delta {m^\prime}^2\vert
\la 10^{-1}$ eV$^2$, the subsequent \mus resonance occurs during the
chaotic phase of the $\nu_\tau\leftrightarrow\nu_{s^\prime}$ evolution.
Our numerical calculations show that
in this case the two mixing systems are coupled and continue to be
chaotic.  The \mus transformation will in fact soon dominate the 
chaotic oscillation process
because this transformation channel has a larger $\vert V_x\vert$
than the BBN-constrained $\nu_\tau\leftrightarrow\nu_{s^\prime}$
transformation channel.  Therefore, for
$10^{-2}$ eV$^2\la\vert\delta {m^\prime}^2\vert\la 10^{-1}$ eV$^2$,
the \mus transformation cannot be suppressed by the presence of
the $\nu_\tau\leftrightarrow\nu_{s^\prime}$ transformation channel.

The $\nu_\tau\leftrightarrow\nu_{s^\prime}$ system is also bounded
by BBN at large $\delta {m^\prime}^2$.
If a large ($L_{\nu_\tau}\ga 0.1$)
asymmetry is generated by the $\nu_\tau\leftrightarrow\nu_{s^\prime}$
transformation before $T\sim 5$ MeV, the produced $\nu_{s^\prime}$
or $\bar\nu_{s^\prime}$ and the re-thermalized $\nu_\tau\bar\nu_\tau$
will increase $N_\nu$ by 0.5, violating the BBN bound.
This imposes an upper bound on $\vert\delta {m^\prime}^2\vert$
\cite{Foot3,Shi3}:
\begin{equation}
\vert\delta {m^\prime}^2\vert\la 10^4\,{\rm eV}^2,
\end{equation}
implying $m_{\nu_\tau}\la 100$ eV if $\nu_{s^\prime}$ is
much lighter.

In figure 2 we plot the conditions that $\nu_\tau$ and
the $\nu_\tau\leftrightarrow\nu_{s^\prime}$ mixing must satisfy
in order to alleviate the BBN constraint on the maximal
vacuum $\nu_\mu\leftrightarrow\nu_s$ mixing that may be 
implied by the SuperK data.

We finally very briefly comment on another scenario in which a positive
lepton asymmetry generated by a $\nu_\tau\leftrightarrow\nu_{s^\prime}$ mixing
may be partially converted into the $\nu_e\bar\nu_e$ sector by a
matter-enhanced $\nu_\tau\leftrightarrow\nu_e$ transformation.
This positive $\nu_e$ chemical potential
can reduce the effective $N_\nu$ and may make some room for extra neutrino
energy density.  A previous paper has argued that the reduction in 
$N_\nu$ can be as large as $\Delta N_\nu\approx 0.5$\cite{Foot3}.
This will not accommodate for an extra $\Delta N_\nu=0.9$ produced
by the \mus  mixing in question, given the $^4$He-derived
bound $N_\nu\la 3.3$.
We also note that even the 0.5 reduction in $N_\nu$ may be
overestimated\cite{Shi3}.

\section{Summary}
We have shown that based on the observationally-inferred
primordial abundance of $^4$He ($Y<0.25$) and
deuterium (D/H$\approx 3.4\pm 0.3\times 10^{-5}$),
Big Bang Nucleosynthesis yields a stringent bound on the effective number
of neutrino species (energy densities) 
during the BBN epoch, $N_\nu\la 3.3$.  This bound
can be employed to constrain active-sterile neutrino mixings, as plotted
in Fig.~1.  With the preassumption that all three active neutrinos are light
compared to 1 MeV, it rules out the
$\nu_\mu\leftrightarrow\nu_s$ mixing explanation of the
Super Kamiokande atmospheric neutrino data. The only way to circumvent
this bound is if there is a simultaneous mixing between
$\nu_\tau$ and a lighter sterile neutrino $\nu_{s^\prime}$
with the following properties:
\begin{enumerate}
\item A mixing mass-squared difference
$200$ eV$^2\la m^2_{\nu_\tau}-m^2_{\nu_{s^\prime}}\la 10^4$ eV$^2$
and a non-radiative $\nu_\tau$ decay lifetime $\tau_{\nu_\tau}\la 10^3$ year.
The mass-squared difference implies
a $\nu_\tau$ mass between $\sim 15$ eV and $\sim 100$
eV if $\nu_{s^\prime}$ is much lighter than $\nu_\tau$.
\item $[(m^2_{\nu_\tau}-m^2_{\nu_{s^\prime}})/{\rm eV}^2]^{1/6}
\sin^22\theta\ga 10^{-11}$ and 
$[(m^2_{\nu_\tau}-m^2_{\nu_{s^\prime}})/{\rm eV}^2]
\sin^42\theta\la 10^{-7}$, where $\theta$ is the vacuum mixing
angle between $\nu_\tau$ and $\nu_{s^\prime}$.
\end{enumerate}
We conclude that the
$\nu_\mu\leftrightarrow\nu_\tau$ vacuum oscillation channel
with maximal or near maximal mixing angle provides a more natural
explanation for the Super Kamiokande data.

We thank Robert Foot and Ray Volkas for helpful discussions.
X.~S. and G.~M.~F. are partially supported by NASA grant NAG5-3062
and NSF grant PHY98-00980 at UCSD.

\newpage
\section{Appendix}

Our semi-analytical calculations track the movement (in energy
space) of resonances in the neutrino energy spectrum and the 
resonant conversion rate of neutrinos in the resonance regions.
For $\nu_\alpha\leftrightarrow\nu_s$ mixing with vacuum mixing 
parameters $\delta m^2=m_{\nu_s}^2-m_{\nu_\alpha}^2<0$ and $\sin^22\theta$,
both the conversion of $\nu_\alpha$ and $\bar\nu_\alpha$ are calculated.
If we denote $\epsilon_{\rm res}^{(\pm)}$ as the resonant neutrino energy
(divided by the temperature) for $\nu_\alpha$ (\lq\lq $+$\rq\rq)
and $\bar\nu_\alpha$ (\lq\lq $-$\rq\rq),
The rate of resonant conversion, when normalized by the total $\nu_\alpha$
or $\bar\nu_\alpha$ number density, is
\begin{equation}
{\cal R}^{(\pm)}\approx \pi f[\epsilon_{\rm res}^{(\pm)}]
\left\vert {\partial\epsilon_{\rm res}^{(\pm)}\over
\partial V_z^{(\pm)}}\right\vert {V_x}^2
=\pi f[\epsilon_{\rm res}^{(\pm)}]
{\vert\delta m^2\vert\sin^22\theta\over 4T}.
\label{convertrate}
\end{equation}
This equation is essentially the l.h.s. of Eq.~(\ref{requirement})
except for a factor $\pi$ which comes from a detailed integration 
of transition probability over the entire resonance region.
The rate of change of $L_{\nu_\alpha}$ is therefore
\begin{equation}
{{\rm d}L_{\nu_\alpha}\over {\rm d}t}={3\over 8}
\left({\cal R}^{(+)}-{\cal R}^{(-)}\right).
\label{growthrate}
\end{equation}

When $L_{\nu_\alpha}$ is small (i.e., $\vert V_\alpha^L\vert\ll
\vert V_\alpha^T\vert$), the effect of matter-antimatter asymmetry on the
$\nu_\alpha\leftrightarrow\nu_s$
and $\bar\nu_\alpha\leftrightarrow\bar\nu_s$ oscillation is small. 
Both oscillation systems will encounter resonances, at an energy
\begin{equation}
\epsilon_{\rm res}^{(\pm)}\approx
\left({\vert\delta m^2\vert\over 44G_F^2T^6}\right)^{1/2}
\left[1\pm {0.35G_FT^3(2L_{\nu_\alpha}+L_0)
\over\vert\delta m^2\vert/2T}\right],
\end{equation}
(see Eq.~[\ref{Vcomponent}] to [\ref{VT}]).
Eq.~(\ref{growthrate}) is approximately
\begin{equation}
{{\rm d}L_{\nu_\alpha}\over {\rm d}t}\approx {\pi\over 4}
{{\rm d}f(\epsilon)\over {\rm d}\epsilon}
G_FT^3\sin^22\theta(L_{\nu_\alpha}+L_0/2),
\end{equation}
with ${\rm d}f(\epsilon)/{\rm d}\epsilon$ 
evaluated at $\epsilon=\vert\delta m^2/44G_F^2T^6\vert^{1/2}$.
This is apparently a damping equation for $(L_{\nu_\alpha}+L_0/2)$
if ${\rm d}f(\epsilon)/{\rm d}\epsilon<0$ (when $\epsilon<2.217$),
and a growth equation for $(L_{\nu_\alpha}+L_0/2)$ if
${\rm d}f(\epsilon)/{\rm d}\epsilon>0$ (when $\epsilon>2.217$).
Therefore, $(L_{\nu_\alpha}+L_0/2)$ is always damped to 0 at temperatures
$T>T^\prime_{\rm res}$, when $\epsilon_{\rm res}^{(\pm)}$ is small;
and $(L_{\nu_\alpha}+L_0/2)$ grows for $T<T^\prime_{\rm res}$ when
$\epsilon_{\rm res}^{(\pm)}$ is large enough.

In Figure 3 we plot the result of our numerical calculation
based on Eq.~(\ref{convertrate}) and (\ref{growthrate}) for 
$\nu_\tau\leftrightarrow\nu_{s^\prime}$ mixing, assuming
$m^2_{\nu_\tau}-m^2_{\nu_{s^\prime}}=50$ eV$^2$ and 
$\sin^2 2\theta^\prime=10^{-4}$.  It is for the most part similar
to the thick solid line in Figure 1 of Foot\cite{Foot4}, which
employs the same parameters.  There are minor differences that are
readily identifiable: (1) our result tracks $T^{-4}$ more closely
in the \lq\lq power-law growth\rq\rq epoch; (2) $L_{\nu_\tau}$ in
our results does not switch sign at the initial point of growth. The
sign difference is not surprising because of the chaotic character of
the growth, which introduces a sign ambiguity to the problem \cite{Shi2}.
Similar calculations for other choices of 
$m^2_{\nu_\tau}-m^2_{\nu_{s^\prime}}$
and $\sin^2 2\theta^\prime=10^{-4}$ show that the temperature
at which the $L_{\nu_\tau}$ growth starts is approximately
given by Eq.~(\ref{Tres22}).

As an illustration, also plotted in Figure 3, are
the $\vert L_{\nu_\tau}\vert$ required for the
$\nu_\mu\leftrightarrow\nu_s$ or $\bar\nu_\mu\leftrightarrow\bar\nu_s$
resonance to occur at $\nu_\mu$ or $\bar\nu_\mu$
energies $\epsilon_{\rm res}\equiv p^{(\mu)}_{\rm res}/T
=0.01,\,1,\,10$.  (The parameters for the
$\nu_\mu\leftrightarrow\nu_s$ mixing are
$\delta m^2=10^{-3}$ eV$^2$ and $\sin 2\theta=1$.)
It can be seen from the figure that the lower
energy component of the $\nu_\mu$ or $\bar\nu_\mu$
neutrinos encounters the resonance first when
$\vert L_{\nu_\tau}\vert$ is very small, and
the resonance region moves through the $\nu_\mu$
or $\bar\nu_\mu$ spectrum to higher neutrino
energies as $\vert L_{\nu_\tau}\vert$ becomes larger.
Essentially all $\nu_\mu$ or $\bar\nu_\mu$ encounter
resonances at $T\approx T^\prime_{\rm res}$.

In Figure 4 we show $\vert {\rm d}\ln L_{\nu_\tau}/{\rm d}\ln T \vert$
as a function of the $\nu_\tau\leftrightarrow\nu_{s^\prime}$
vacuum mixing parameters, $m^2_{\nu_\tau}-m^2_{\nu_s}$
and $\sin^22\theta^\prime$, in the initial exponential stage of
$L_{\nu_\tau}$ growth when $L_{\nu_\tau}$ is $\ll 10^{-7}$.
This figure shows that
$\vert {\rm d}\ln L_{\nu_\tau}/{\rm d}\ln T \vert$
is approximately a linear function of $\sin2\theta^\prime$,
and is insensitive to $m^2_{\nu_\tau}-m^2_{\nu_s}$.
An analytical fit to this numerical result yields Eq.~(\ref{upperbound}).

\newpage
\noindent{\bf Figure Captions:}

\noindent
Figure 1. $^4$He yields of the 
$\nu_\mu\,({\rm or}\,\nu_\tau)\leftrightarrow\nu_s$ mixing
(for $\eta=4.7\times 10^{-10}$).  Mixings that yield
$Y\ge 0.25$, including the region inside the dashed line that
is compatible with the Super Kamiokande data, are ruled out by BBN.
\bigskip

\noindent
Figure 2. The parameter space of 
$\nu_\tau\leftrightarrow\nu_{s^\prime}$ mixing and
the additional requirements on $\nu_\tau$ that may
alleviate the BBN constraint on the
$\nu_\mu\leftrightarrow\nu_s$ mixing parameters
which can accomodate the Super Kamiokande data.
\bigskip

\noindent
Figure 3. The growth of the tau neutrino asymmetry as
a result of the tau neutrino-sterile neutrino mixing, assuming
$m^2_{\nu_\tau}-m^2_{\nu_{s^\prime}}=50$ eV$^2$ and
$\sin^2 2\theta^\prime=10^{-4}$.
The intersections between the growth curve for $L_{\nu_\tau}$
and the dashed lines indicate when resonances occur for
$\nu_\mu$ (if $L_{\nu_\tau}>0$) or $\bar\nu_\mu$ (if $L_{\nu_\tau}<0$)
neutrinos with momentum $p$.
\bigskip

\noindent
Figure 4. The initial rate for $L_{\nu_\tau}$ growth,
${\rm d}\ln L/{\rm d}\ln T$, as a function of the
vacuum tau neutrino-sterile neutrino mixing parameters.
\bigskip


\newpage
\begin{references}
\bibitem{SuperK} Y.~Fukuda, {\sl et al.}, {\sl Phys. Rev. Lett.} {\bf 81},
		 1562 (1998).

\bibitem{sterile} M.~C.~Gonzalez-Garcia, H.~Nunokawa, O.~L.~G.~Peres,
                 T.~Stanev, J.~W.~F.~Valle, {\sl Nucl. Phys.} {\bf B},
                 in press (1998).

\bibitem{Solar} See e.g., N.~Hata and P.~G.~Langacker, {\sl Phys. Rev.}
	       {\bf D 56}, 6107 (1997).

\bibitem{Dolgov} R.~Barbieri and A.~Dolgov, {\sl Phys. Lett.} {\bf B 349},
		 743 (1991).

\bibitem{Enqvist1} K.~Enqvist, K.~Kainulainen, and M.~Thomson,
	           {\sl Nucl. Phys.} {\bf B 373}, 498 (1992).

\bibitem{Cline} J.~M.~Cline, {\sl Phys. Rev. Lett.}, {\bf 68}, 3137 (1992).

\bibitem{Shi1} X.~Shi, D.~N.~Schramm, and B.~D.~Fields, {\sl Phys. Rev.}
	       {\bf D 48}, 2568 (1993).

\bibitem{Foot1} R.~Foot, M.~J.~Thomson, and R.~R.~Volkas,
		{\sl Phys. Rev.} {\bf D 53}, 5349 (1996).

\bibitem{Shi2} X.~Shi, {\sl Phys. Rev.} {\bf D 54}, 2753 (1996).

\bibitem{Foot3} R.~Foot and R.~R.~Volkas, {\sl Phys. Rev.} {\bf D 56},
		6653 (1997).

\bibitem{Raffelt} D.~N\"otzold and G.~Raffelt, 
	           {\sl Nucl. Phys.} {\bf B 307}, 924 (1988).

\bibitem{McKeller} B.~H.~J.~McKeller and M.~J.~Thomson,
		   {\sl Phys. Rev.} {\bf D 49}, 2710 (1994).

\bibitem{Malaney} G.~M.~Fuller and R.~A.~Malaney, {\sl Phys. Rev.}
                  {\bf D 43}, 3136 (1991).

\bibitem{Schramm} G.~Steigman, D.~N.~Schramm and J.~E.Gunn, {\sl Phys. Lett}
	          {\bf B 66}, 262 (1977); for a recent update, see
  		  C.~Y.~Cardall and G.~M.~Fuller, {\sl Astrophys. J.}
		  {\bf 472}, 435 (1996).

\bibitem{Tytler1} S.~Burles and D.~Tytler, to appear in the {\sl
	Proceedings of the Second Oak Ridge Symposium on Atomic \& Nuclear
	Astrophysics}, ed. A. Mezzacappa (Institute of Physics, Bristol),
	and references therein.

\bibitem{Cardall} C.~Y.~Cardall and G.~M.~Fuller, {\sl Phys. Rev.}
		  {\bf D 54}, 1260 (1996).

\bibitem{Olive} K.~A.~Olive, E.~Skillman, and G.~Steigman,
		{\sl Astrophys. J.}, {\bf 483}, 788 (1997).

\bibitem{Izotov} Y.~Izotov and T.~X.~Thuan, {\sl Astrophys. J.}, {\bf 500},
		 188 (1998).

\bibitem{Dorman} B.~Dorman, Y.~Lee, and D.~A.~VandenBerg,
                        {\sl Astrophys.~J.} {\bf 366}, 115 (1991).

\bibitem{Foot2} R.~Foot and R.~R.~Volkas, {\sl Phys. Rev.} {\bf D 55},
		5147 (1997).

\bibitem{Foot4} R.~Foot, unpublished (1998). 


\bibitem{Kolb} E.~W.~Kolb and M.~S.~Turner, {\sl The Early Universe},
	       (Addison-Wesley, Reading, MA, 1990).

\bibitem{Kauffmann} G.~Kauffmann and S.~Charlot,
                    {\sl Astrophys.~J.}, {\bf 430}, L97 (1994).

\bibitem{Klypin} A.~Klypin, S.~Borgani, J.~Holtzman, and J.~Primack,
                    {\sl Astrophys.~J.}, {\bf 444}, 1 (1995).

\bibitem{Ma97} C.~Ma {\sl et al.}, {\sl Astrophys.~J.}, {\bf 484}, L1 (1997).

\bibitem{Gross} See e.g.,
		M.~A.~K.~Gross {\sl et al.}, {\sl Mon. Not. R. Astron. Soc.},
	        in press (1998).

\bibitem{Kirshner} A.~G.~Riess {\sl et al.}, {\sl Astrophys.~J.},
		   in press (1998).

\bibitem{Raffelt2} G.~G.~Raffelt,  {\sl Astrophys.~J.} {\bf 365},
		  559 (1990), and references therein.

\bibitem{Shi3} X.~Shi, G.~M.~Fuller and K.~Abazajian (in preparation). 

\end{references}
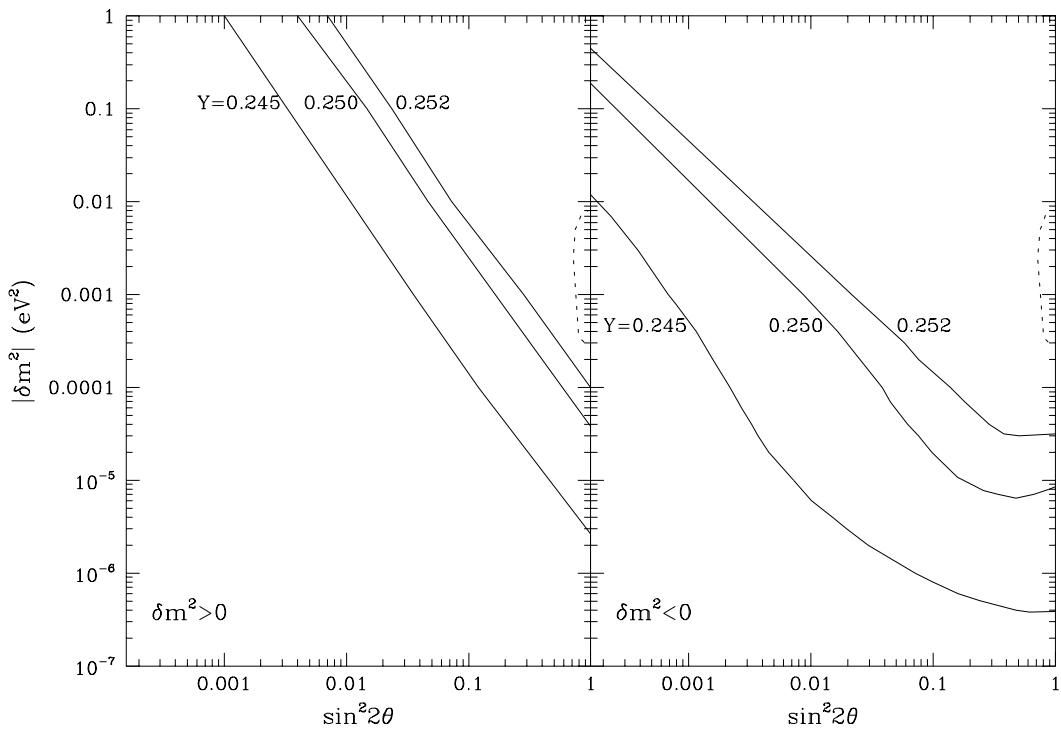
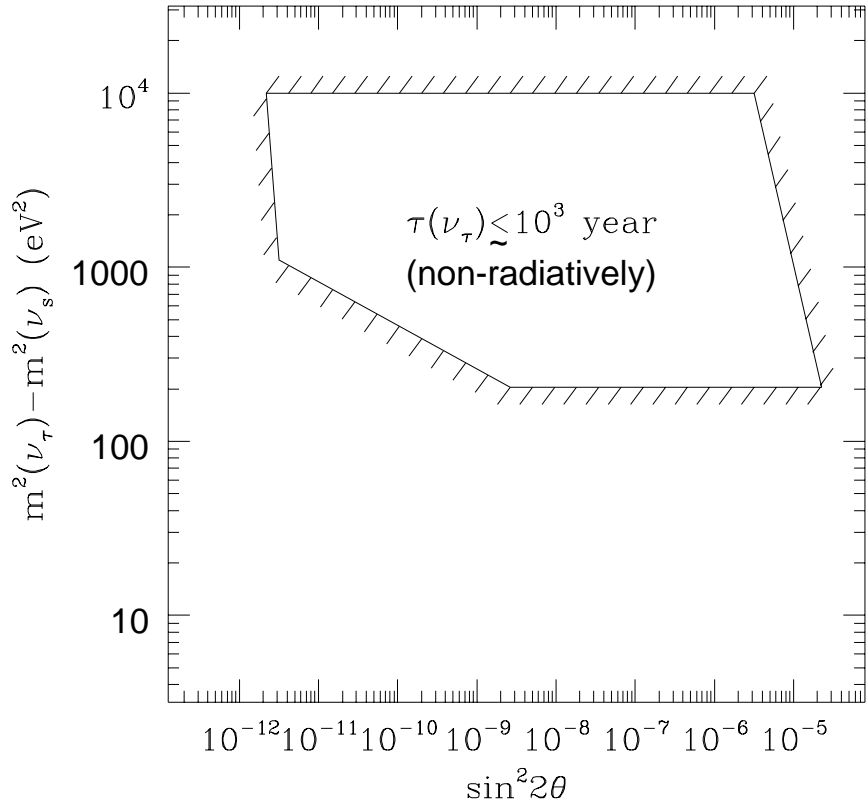
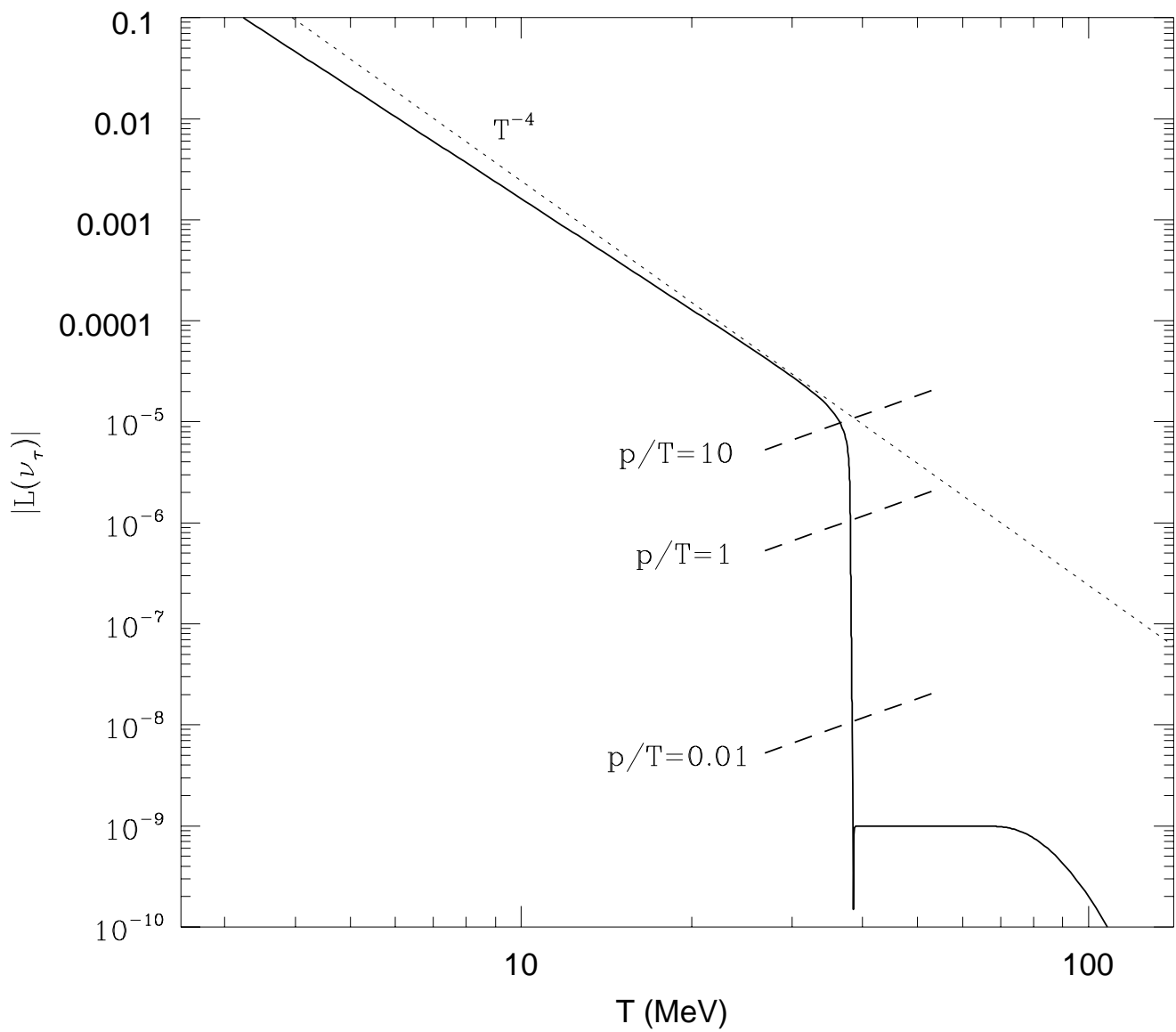
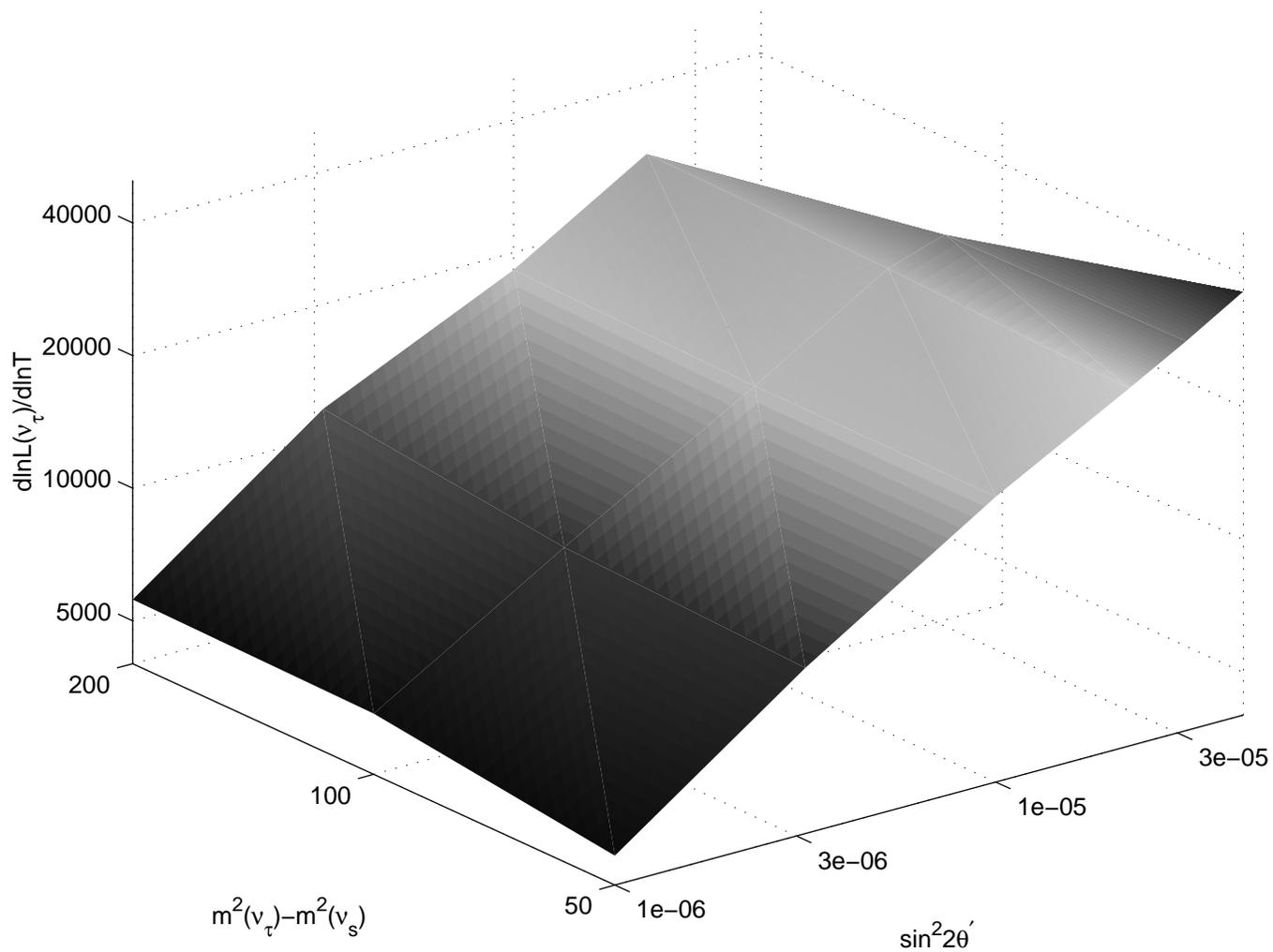
\end{document}